\documentclass[12pt]{iopart}

\usepackage{graphicx}
\usepackage{iopams}
\usepackage{bm}

\begin{document}
\title{Spin fragmentation of Bose-Einstein condensates with antiferromagnetic interactions}

\author{Luigi De Sarlo$^1$ \footnote{Current address : SYRTE, Observatoire de Paris, CNRS, UPMC ; 
61 avenue de l'Observatoire, 75014 Paris, France}, Lingxuan Shao$^1$ , Vincent Corre$^1$ , Tilman Zibold$^1$ , David Jacob$^1$ , Jean Dalibard$^1$  and Fabrice Gerbier$^1$ }

\address{$1$  Laboratoire Kastler Brossel, CNRS, ENS, UPMC, 24 rue Lhomond, 75005 Paris}
\ead{fabrice.gerbier@lkb.ens.fr}

\date{\today}

\begin{abstract}
We study spin fragmentation of an antiferromagnetic spin 1 condensate in the presence of a quadratic Zeeman (QZ) effect breaking spin rotational symmetry. We describe how the QZ effect turns a fragmented spin state, with large fluctuations of the Zeemans populations, into a regular polar condensate, where atoms all condense in the $m=0$ state along the field direction.  We calculate the average value and variance of the Zeeman state $m=0$ to illustrate clearly the crossover from a fragmented to an unfragmented state. The typical width of this crossover is $q \sim k_B T/N$, where $q$ is the QZ energy, $T$ the spin temperature and $N$ the atom number. This shows that spin fluctuations are a mesoscopic effect that will not survive in the thermodynamic limit $N\rightarrow \infty$, but are  observable for sufficiently small atom number.     
\end{abstract}
\maketitle
\section{Introduction}\label{int}

The natural behavior of bosons at low enough temperatures is to form a Bose-Einstein condensate, {\it i.e.} a many-body state where one single-particle state becomes macroscopically occupied \cite{pitaevskii2003a}. There are, however, situations where bosons can condense simultaneously in several single-particle states, forming a so-called {\emph{fragmented condensate}}. Several examples are known, where fragmentation occurs due to orbital (Bose gases in optical lattices or in fast rotation) or to internal degeneracies (pseudo-spin $1/2$ or spin $1$ Bose gases). These examples have been reviewed in \cite{castin2001a,mueller2006a}. 

The spin 1 Bose gas, first studied by Nozi\`eres and Saint James \cite{nozieres1982a}, is a striking example where fragmentation occurs due to rotational symmetry in spin space. For antiferromagnetic interactions of the form $V_{12}=g_s {\bf s}_1 \cdot {\bf s}_2$ between two atoms with spins ${\bf s}_{1}$ and ${\bf s}_{2}$ ($g_s>0$), the many-body ground state is expected to be a spin singlet state \cite{castin2001a,law1998a}. In such a state the three Zeeman sublevels are occupied, leading to three macroscopic eigenvalues of the single-particle density matrix (instead of just one for a regular condensate). As pointed out in \cite{law1998a,ho2000a,castin2001a}, the signature of fragmentation is then the occurrence of anomalously large fluctuations of the populations $N_m$ in the Zeeman states $m=0,\pm 1$ (see also \cite{kuklov2002a}, where a similar behavior is predicted in a pseudo spin $1/2$ system). In the singlet state for instance, the expectation value and variance of $N_0$ are $\langle N_{0}\rangle = N/3$, and $ \Delta N_{0}^2 \approx4N^2/45$, respectively ($N$ is the total number of particles). Such super-Poissonian fluctuations ($\Delta N_{0}^2\propto \langle N_{0}\rangle ^2$) deviate strongly from the value expected for a single condensate or any ensemble without correlations where $\Delta N_{0}^2\propto \langle N_{0}\rangle$  \footnote{Note that the problem we discuss here is unrelated to the anomalous fluctuations of the {\it total} condensate number found for ideal gases in the grand canonical ensemble \cite{pitaevskii2003a}. In this work, we assume implicitly the canonical ensemble, and study the fluctuations of the populations of individual Zeeman states discarding quantum and thermal depletion of the condensate. }.  It was pointed out by Ho and Yip \cite{ho2000a} that such state was likely not realized in typical experiments, due to its fragility towards any perturbation breaking spin rotational symmetry (see also \cite{zhou2001a,cui2008a,barnett2010a,barnett2011a,lamacraft2011a,tasaki2013a}). In the thermodynamic limit $N\rightarrow \infty$, an arbitrary small symmetry-breaking perturbation is enough to favor a regular condensed state, where almost all atoms occupy the same (spinor) condensate wave function and $\Delta N_{0} \ll N$. 

In this article, we give a detailed analysis of the phenomenon of spin fragmentation for spin 1 bosons. Our analysis assumes the conservation of the total magnetization $m_z$. The fact that magnetization is an (almost) conserved quantity follows from the rotational invariance of the microscopic spin exchange interaction, and from the isolation of atomic quantum gases from their environment. 
A key consequence is that in an external magnetic field $B$, the linear Zeeman effect only acts as an energy offset and does not play a role in determining the equilibrium state. The dominant effect of an applied magnetic field is a second-order (or quadratic) Zeeman energy, of the form $q (m^2-1)$ for a single atom in the Zeeman state with magnetic quantum number $m$ \footnote{This second-order shift originates from the hyperfine coupling between electronic and nuclear spins, and corresponds to the second order term in an expansion of the well-known Breit-Rabi formula for alkalis (sees, {\it e.g.}, \cite{foot_atomic}).}.  The QZ energy breaks the spin rotational symmetry, and favors a condensed state with $m=0$ along the field direction. In \cite{cui2008a,barnett2010a,barnett2011a,lamacraft2011a}, the evolution of the  ground state with the QZ energy $q$ was studied theoretically. Since experiments are likely to operate far from the ground state, it is important to understand  quantitatively how the system behaves at finite temperatures. This is the main topic we address in this paper. 

\begin{figure}[ht!!!!]
\begin{center}
\includegraphics[width=12cm]{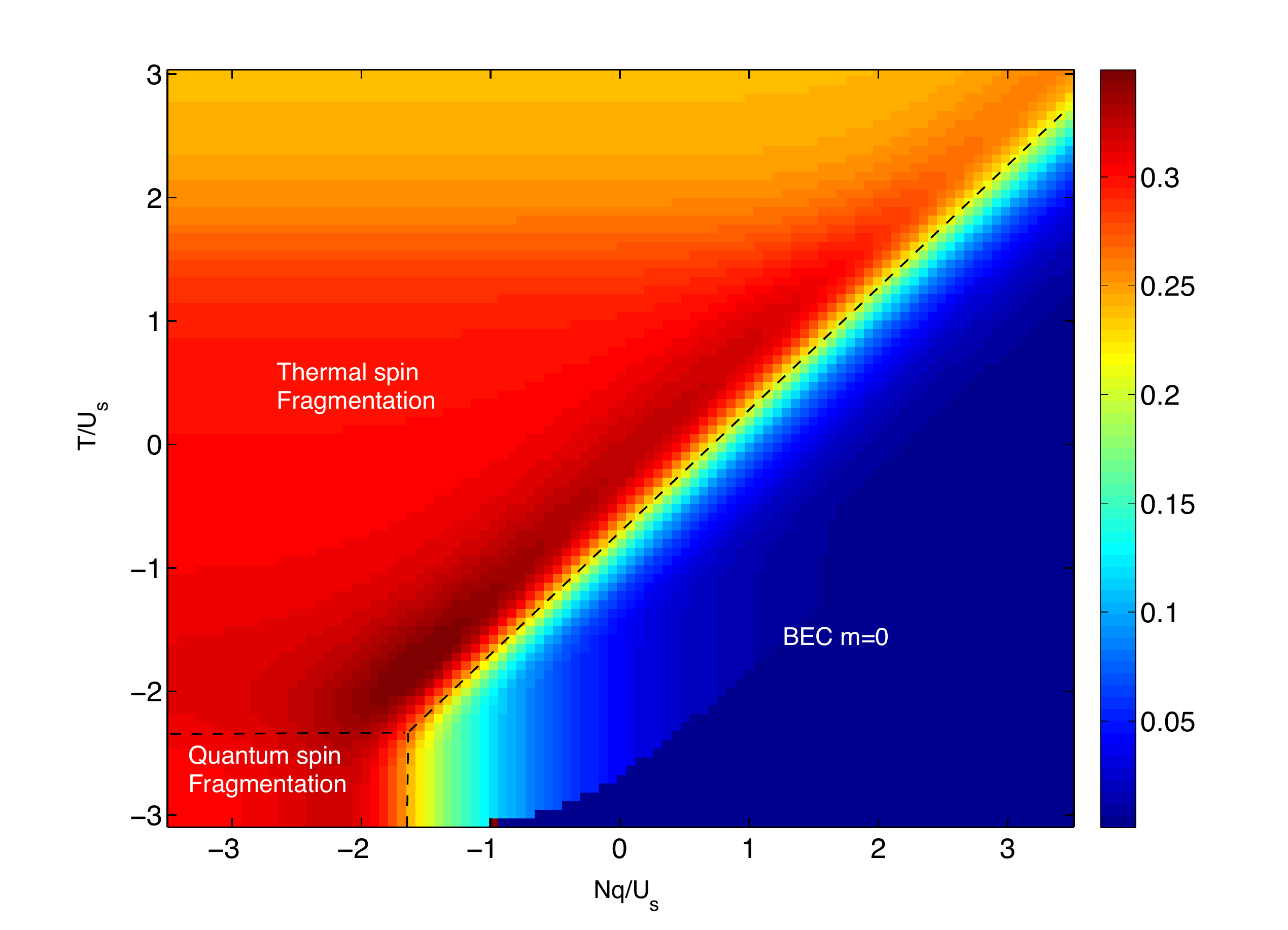}
\end{center}
\caption{Standard deviation $\Delta n_0=\Delta N_0/N$ of the population $N_0$ of the Zeeman state $m=0$, normalized to the total atom number $N$. We mark three different regimes in the $q-T$ plane. ``Spin fragmentation'' refers to a fragmented spin state with large population fluctuations, where $\Delta n_0\sim 1$. In the quantum regime ($Nq/U_s\ll 1/N$ and $k_B T/U_s\ll 1/N$), this is due to quantum fluctuations: the system is then close to the singlet ground state. In the thermal regime($k_B T\gg Nq,U_s/N$), on the other hand, thermal fluctuations dominate over the quantum one and over the effect of the QZ energy (QZ energy). Conversely, "BEC m=0" refers to atoms forming a regular polar condensate with almost all atoms in $m=0$, and $\Delta n_0\ll 1$. The plot was drawn by numerically diagonalizing the Hamiltonian (\ref{eq:H}) and computing thermodynamic averages from the spectrum and eigenstates, using $N=300$ particles. Note the logarithmic scales on both the horizontal and the vertical axis. }
\label{Fig:summary}
\end{figure}

Our focus in this article will be to calculate the first two moments (average value and variance) of $N_0$. These moments illustrate clearly the evolution of the system from fragmented to unfragmented and thus constitute the main experimental signature of fragmentation. The main findings are summarized in Figure~\ref{Fig:summary}, where we plot the standard deviation of $n_0=N_0/N$ in a $q-T$ plane. Large fluctuations and depletion of the $m=0$ state are observed for small $q$. We can distinguish three different regimes. For low $q \ll U_s/N^2$ and low temperatures $k_B T \ll U_s/N$ ($U_s\propto g_s$ is the spin interaction energy per atom), the system is close to the ground state in a regime we call ``quantum spin fragmented" \cite{law1998a,ho2000a,castin2001a,barnett2011a}. We also observe a thermal regime for $ k_B T \gg Nq,U_s/N$ dominated by thermally populated excited states. We call this second regime ``thermal spin fragmented''. Finally, for $q$ large enough and temperature low enough, the bosons condense into the single-particle state $m=0$, forming a so-called ``polar'' condensate \cite{ho1998a,ohmi1998a}. In this limit, $\langle N_0 \rangle \approx N$ and $\Delta N_0 \ll N$.  We indicate this third regime as ``BEC$ m=0$'' in Figure~\ref{Fig:summary}. 

The evolution from the fragmented, singlet condensate to an unfragmented condensate with increasing QZ energy $q$ is similar to a well-known example in the literature on quantum magnetism, the Lieb-Matthis model of lattice Heisenberg antiferromagnets \cite{lieb1962a}. This model describes collective spin fluctuations of an Heisenberg antiferromagnet on a bipartite lattice. It constitutes a popular toy model for demonstrating the appearance of broken symmetry ground states in condensed matter \cite{kaiser1989a,kaplan1990a,bernu1992a,vanwezel2005a,vanwezel2006a}. The ground state of such system (in principle also a spin singlet) was found theoretically to evolve to a N\'eel state in the thermodynamic limit in the presence of an arbitrarily small {\it staggered } magnetic field (whose sign alternates from one site to the next). The underlying theory is close to the one presented here. An essential difference is that the present model of antiferromagnetic spin 1 BECs is expected to accurately describe actual experimental systems \cite{black2007a,jacob2012a}. In the antiferromagnet case, the staggered magnetic field is a theoretical object that cannot be produced in the laboratory for real solids. In contrast, the QZ energy is easily controllable in spin 1 BEC experiments. Another important difference is that experiments with ultracold quantum gases are typically done with relatively small atom numbers, from $N\sim 10^2$ to $N\sim 10^6$, so that conclusions that hold in the thermodynamic limit do not necessarily apply and spin fragmentation can  be observed experimentally. 

The article is organized as follows. In Section \ref{sec:model}, we present the basic model that describes an ensemble of spin 1 bosons with antiferromagnetic interactions condensing in the same orbital wave function  irrespective of the internal state (single-mode approximation, or SMA). In Section \ref{sec:spectrum}, we use the basis of total spin eigenstates (exact in the absence of an applied magnetic field, $q=0$). We derive approximate solutions for the spectrum and eigenstates for $q>0$ in section \ref{sec:TB}, and discuss how they evolve with increasing QZ energy. Using these results, we compute in section \ref{sec:SMAT} the average value and variance of $N_0$ at finite temperatures, and compare the approximate solution to numerical diagonalization of the Hamiltonian. We finally present in Section \ref{sec:broken} an alternative approach, where the fragmented condensate is described as a statistical mixture of mean-field (symmetry broken) states. We find excellent agreement with the exact diagonalization of the Hamiltonian. 

\section{Single-mode description of spin 1 condensates}
\label{sec:model}

We consider a gas of ultracold spin 1 bosons in a trap with Zeeman components $m=-1,0$,or $+1$. We discuss the case of antiferromagnetic interactions and assume the validity of the SMA, {\it i.e.} that all bosons condense in the same spatial orbital irrespective of their internal state \cite{yi2002a}. The Hamiltonian is \cite{stamperkurn2012a}
\begin{equation}
\label{eq:H}
\hat{H} = \frac{U_s}{2N} {\bf \hat S}^2 - q \hat{N}_{0},
\end{equation}
where $U_s>0$ is the spin interaction energy per atom \footnote{The spin interaction energy $U_s$ can be calculated from $U_s = g_s \int d \mathbf{x} |\psi(\mathbf{x})|^4$, where $\psi(\mathbf{x})$ is the spatial orbital of the condensate.}, $q>0$ is the QZ energy, $ {\bf \hat S}$ is the total spin operator, and $\hat{N}_{\alpha}$ is the number operator in the Zeeman state $\alpha$. We assume that the number of atoms $N$ is even for simplicity. Odd values of $N$ could be treated in a similar way, without modifying the final results to order $1/N$. Typical experimental values for the parameters of the SMA model are $N=10^3-10^5$, $U_s/k_B \sim 2-5~$nK, while $q$ can be varied from zero to values much larger than $U_s$ by changing the magnetic field \cite{black2007a,jacob2012a}.



In the absence of an external magnetic field ($q=0$), the Hamiltonian reduces to a quantum rotor with moment of inertia $N/U_s$ \cite{law1998a,barnett2010a}. The energy eigenstates are thus simply the total spin eigenstates $\vert N,S,M \rangle$, with $S$ the spin quantum number and $M$ its projection on the $z$ axis. The corresponding eigenvalues are $E(S)= (U_s/2N)S(S+1)$, with a degeneracy $2S+1$. The wave functions for these states are known explicitly in the Fock basis \cite{law1998a,ho2000a,castin2001a} (see also \ref{app:wf}). 

When $q\neq 0$,  since $[\hat{S}_z,\hat{N}_{0}]=0$,  the magnetic quantum number  $M$ (eigenvalue of $\hat{S}_z$) remains a good quantum number. One can diagonalize $\hat{H}$ by block in each $M$ sector. For each $M$, the energy eigenstates can be expressed in the angular momentum basis,
\begin{eqnarray}
\vert \phi_M \rangle = \sum_{S=\vert M\vert}^{N}  c_{S,M} \vert N,S,M \rangle.
\end{eqnarray}
To express the Hamiltonian in (\ref{eq:H}) in the $\vert N,S,M \rangle$ basis, we need to compute the action of $\hat{N}_0$. The non-vanishing matrix elements of $\hat{N}_{0}$ are $
\langle N,S,M \mid \hat{N}_{0}\vert N,S,M\rangle$, $\langle N,S\pm2,M \mid \hat{N}_{0}\vert N,S,M\rangle$ (see \ref{app:wf}). The Schr\"odinger equation then takes the form of a tridiagonal matrix equation,
\begin{eqnarray}
\label{eq:matrix_equation}
h_{S,S+2}^M c_{S+2,M}+h_{S,S-2}^M c_{S-2,M} +h_{S,S}^M  c_{S,M}  =E c_{S,M},
\end{eqnarray}
with $E$ the energy eigenvalue and where the coefficients $h_{S,S'}^M$ are easily obtained from the expressions given in \ref{app:wf}.

\section{Spectrum and eigenstates for $M=0$}

\label{sec:spectrum}

A first approach for finding the spectrum and eigenstates is to diagonalize numerically the matrix $h^{M}$ in (\ref{eq:matrix_equation}). Our goal this Section is to propose an analytical approximation to understand better the structure of the spectrum and eigenstates. The discussion allows one to describe how the ground state evolves with $q$, and will also be useful to understand qualitatively the behavior of the systems at finite temperatures later in this paper. For simplicity, we focus in this Section on the $M=0$ sector. The conclusions we obtain remain qualitatively correct for $M \neq 0$ provided its value is not too large ($\vert M \vert \ll N$).

\subsection{Continuum approximation for large $q$ }
\label{sec:TB}

We make the assumption that the thermodynamic behavior is dominated by states, such that the dominant coefficients in the $\vert N,S,M\rangle$ basis obey $1 \ll S \ll N$. As we will see later in this paper, this assumption is justified for large enough $q$ at $T=0$, and for any $q$ at finite temperatures $k_BT \gg U_s/N$. In this limit, the matrix elements $h_{S,S}, h_{S,S \pm 2}$ can be simplified. We obtain to lowest order in $1/S, S/N$ (see \ref{app:TB}),
\begin{eqnarray}\label{eq:TB}
-J(x+\epsilon) c(x+\epsilon)-J(x-\epsilon)c(x-\epsilon)& +\frac{NU_s}{2} x^2 c(x)=\left(E +\frac{Nq}{2}\right) c(x),
\end{eqnarray}
where we have set $x=S/N$, $\epsilon =2/N$, $c(x) = c_{S,0}$. This equation maps the spin problem to a tight-binding model for a particle hopping on a lattice, with an additional harmonic potential keeping the particle near $x=0$. The model is characterized by an inhomogeneous tunneling parameter $J(x)=Nq(1-x^2/2)/4$ and a harmonic potential strength $NU_s$. Boundary conditions confine the particle to $0 \leq x \leq 1$. 

%
If $c(x)$ changes smoothly as a function of $x$, the tight-binding model can be further simplified in  a continuum approximation. We show in \ref{app:TB} that the tight-binding equation reduces to the one for a fictitious one-dimensional harmonic oscillator,
\begin{eqnarray}\label{eq:HO}
-\frac{q}{N} c''(x)+\frac{N}{4}\left(q+2 U_s\right) x^2 c(x)=\left(E+Nq\right)c(x).
\end{eqnarray}
The boundary condition $c(0)=0$ selects eigenstates  of the standard harmonic oscillator with odd parity. The mass $m$ and oscillation frequency $\omega$ of the fictitious oscillator are found from $\hbar^2/2m \equiv q/N$ and $m\omega^2 \equiv N(q+2U_s)/2$. The oscillator frequency is thus
\begin{eqnarray}\label{eq:bogofreq}
\hbar \omega = \sqrt{q(q+2 U_s)}.
\end{eqnarray}
This collective spectrum was also obtained by the Bogoliubov approach of \cite{cui2008a,barnett2011a}.


\subsection{Ground state}\label{sec:GSHO}

In this Section, we use the results established previously to examine the evolution of the ground state with increasing $q$. Our results reproduce the ones from \cite{barnett2011a} obtained using a different method. The ground state of the truncated  fictitious harmonic oscillator (with boundary condition $c_0(0)=0$) is given by
\begin{eqnarray}
c_0(x) & = \frac{1}{\pi^{1/4}\sigma^{1/2}} \frac{x}{\sigma} \exp\left(-\frac{x^2}{2\sigma^2}\right),
\end{eqnarray}
with the quantum harmonic oscillator size 
\begin{eqnarray}\label{eq:sigma}
\sigma =\sqrt{\frac{\hbar}{m \omega}} = \sqrt{\frac{2}{N}} \left( \frac{q}{q+2U_s} \right)^{1/4}.
\end{eqnarray}
The continuum approximation is valid only if $c(x)$ varies smoothly on the scale of the discretization step $\epsilon$, or equivalently when $\sigma \gg 1/N$. This gives the validity criterion for this approximation,
\begin{eqnarray}
q \gg\frac{U_s}{N^2}
\end{eqnarray}
For $q <U_s/N^2$, the ground state is very close to the singlet state, with a width $\sigma \ll 1/N$. Here spin fragmentation occurs purely due to quantum spin fluctuations (related to antiferromagnetic interactions) of a polar BEC . We indicate this state in Figure~\ref{Fig:summary} as ``quantum spin fragmented".  

For $q \gg U_s/N^2$, the continuum approximation is valid. We see from (\ref{eq:sigma}) that as $q$ increases, the QZ energy mixes an increasing number of $S$ states. Asymptotically, for $q\gg U_s$, the true ground state is a superposition of $\sim N\sigma\approx \sqrt{2N}$ total spin eigenstates.  In this regime, we can compute the moments of $N_0$ by expressing the depletion operator $N-\hat{N}_0$ in terms of the ladder operators $\hat{b}$ and $\hat{b}^\dagger$ associated with the fictitious harmonic oscillator. We find
\begin{eqnarray}
N-\langle N_0\rangle & = \frac{U_s+q}{2\sqrt{q(q+2U_s)}},\\
\Delta N_0^2 & =  \frac{U_s^2}{2q(q+2U_s)}.
\end{eqnarray}
For $U_s/N^2 \ll q \ll U_s$,  the depletion $N-\langle N_0\rangle$  and variance $\Delta N_0^2$ are larger than unity but small compared to $N, N^2$, respectively, while for $q\gg U_s$, they become less than one particle : in the latter case, the ground state approaches the Fock state $\left( \hat{a}_0^\dagger \right)^N \vert \rm vac \rangle$ expected from mean field theory. We indicate both regimes as ``BEC m=0'' in Figure~\ref{Fig:summary}, without marking the distinction. 

\subsection{Excited states for $M=0$}\label{sec:excitedM0}

We now turn to the description of excited states, still limiting ourselves to the case $M=0$ for simplicity. The tight-binding model (\ref{eq:TB}) is characterized by a tunneling parameter $J=Nq(1-x^2/2)/4$ and a harmonic potential strength $\kappa = N U_s$. Let us examine two limiting cases. For $q=0$ (no hopping), the energy eigenstates coincide with ``position'' eigenstates with energy $E(S)\approx (U_s/2N)S^2$ for $S \gg 1$. Conversely, when $U_s=0$ the energy eigenstates are delocalized states, which form an allowed energy band of width $\sim 4J \sim Nq$. The weak inhomogeneity of the tunneling parameter does not play a large role since these states are confined near $x=0$ by the harmonic potential.

\begin{figure}[ht!]
\begin{center}
\includegraphics[width=10cm]{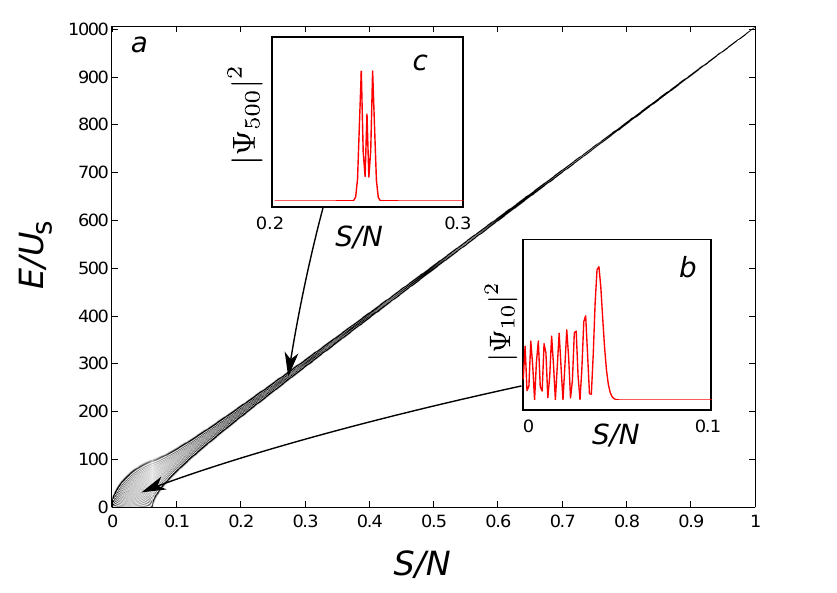}
\caption{Probability densities (amplitude shown as gray scale) of the eigenstates of the spin 1 Hamiltonian (\ref{eq:H}) as a function of ``position'' $x=S/N$ and energy $E/U_s$.  The plot corresponds to $N=2000$ and $Nq/U_s=10$. At low energy, the eigenstates explore the whole available region, from the turning point down to $x=0$. Conversely, at high energies the eigenstates are more and more localized around the diagonal, as expected for potential energy eigenstates. We show as insets the probability densities for the 10th (b) and 500th (c) excited states for illustration.}
\label{Fig:wavefunctions}
\end{center}
\end{figure}
\begin{figure}
\begin{center}
\includegraphics[width=10cm]{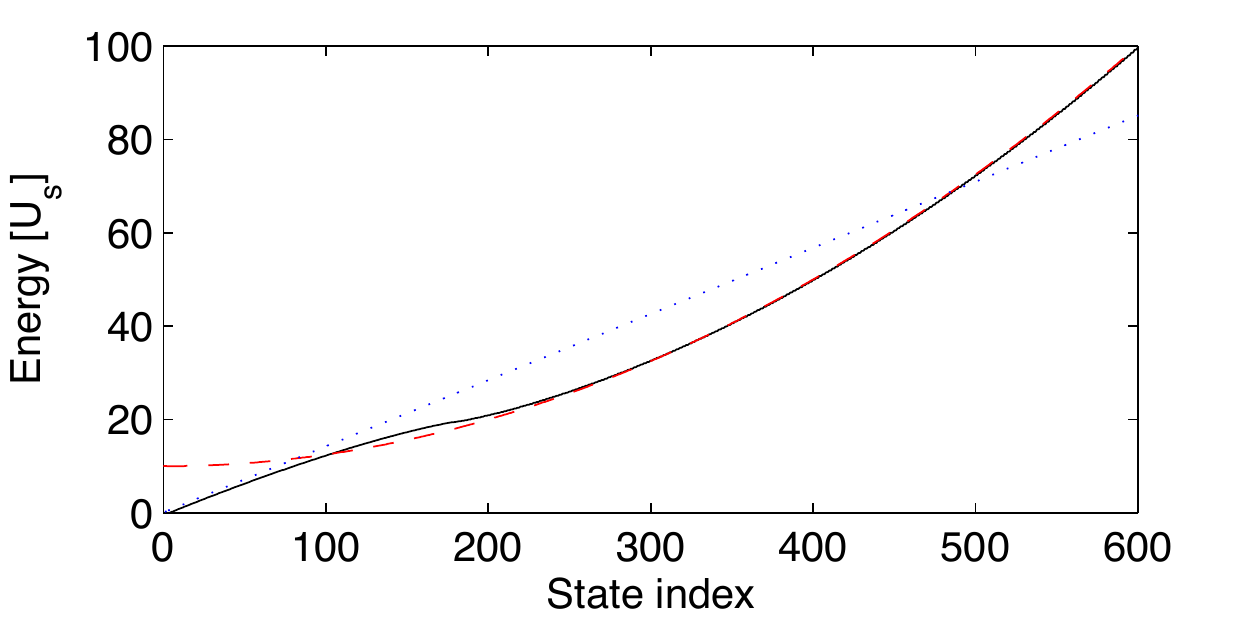}
\caption{Energy spectrum for $N=2000$ and $Nq/U_s=10$. The black solid line is the spectrum calculated by numerical diagonalization of the Hamiltonian (\ref{eq:H}), shifted up by $q N$. The red dashed line corresponds to $E(S)=(U_s/2N)S^2+qN/2$, the blue dotted line to the harmonic oscillator approximation.}
\label{Fig:spectrum}
\end{center}
\end{figure}

For the general case where $J,\kappa\neq0$, the eigenstates can be divided in two groups \cite{hooley2004a,ott2004a},
\begin{itemize}
\item low-energy states with energy $E< 4J$, which are extended ``Bloch-like" states modified by the harmonic potential;   the continuum approximation introduced earlier corresponds to an effective mass approximation, valid for low-energy states with $E \ll 4J=Nq$ (the requirement $q\gg U_s/N^2$ found before still holds). 
\item high-energy states with $E \gg 4J$, that would be in the band gap in the absence of the potential energy term (and thus forbidden).They are better viewed as localized states, peaked around $x(E)\approx \sqrt{2 E/NU_s}$ with a width $\sim1/N$. As a result they are very similar to the angular momentum eigenstates $\vert N,S,0\rangle$ for the corresponding value of $S$. For these states, the continuum approximation does not hold. 
\end{itemize}
We illustrate this classification in Figure~\ref{Fig:wavefunctions}, where we show the probability densities $\mid c(S) \mid^2$ as a function of energy.  One can see  the change from a ``delocalized" regime at small energies to a ``localized" regime at large energies. The wave functions were calculated exactly by diagonalizing the Hamiltonian \ref{eq:H} for $N=1000$. We also show the corresponding energy spectrum in Figure~\ref{Fig:spectrum},  showing  the same crossover from delocalized states at low energies to localized states at high energies. For low energies, the spectrum is given by the harmonic oscillator model, $\epsilon_n\approx \hbar \omega (2n+3/2)$ with $n$ integer. For high energies, the energy eigenstates are localized around $x_n= n/N$, with a spectrum given by $\epsilon_n \approx  U_s n^2/2N$ with $n$ integer. Both expressions agree well with the numerical result in their respective domains of validity.

\section{Spin fragmentation at finite temperatures}

\label{sec:SMAT}
We have seen in Section \ref{sec:GSHO} that for a system in its ground state, the depletion and fluctuations of the $M=0$ state were rapidly collapsing as $q$ was increased above $U_s/N^2$, and the system turned from a fragmented to a single condensate with all atoms in the Zeeman state $m=0$. The energy gap to the first excited state is $3 U_s/N$ near $q=0$. For typical experimental values \cite{black2007a,jacob2012a}, this corresponds to a few pK, vastly smaller than realistic temperatures for a typical experiment (a few tens of nK) due to the $1/N$ scaling. Therefore, it is natural to ask how the crossover from a fragmented to a single condensate is modified at finite temperatures. In the remainder of the paper, we thus consider the high temperature case $k_B T \gg U_s/N$. We will compute the first two moments of $N_0$ at finite temperatures, $\langle N_{0}\rangle_{T}$ and $\left( \Delta N_{0}^2 \right)_{T }=\langle N_{0}^2\rangle_{T}-\langle N_{0}\rangle_{T}^2$, and use these quantities to study the fragmented to single condensate crossover. 

\subsection{Spin fragmentation for $q=0$}
\label{sec:spinfragq0}
Let us first consider the case $q=0$. An important remark is that super-Poissonian fluctuations are not unique to the ground state, but also occur for low-energy eigenstates with $S \ll N$. This is best seen by considering values of $S$ such that $1 \ll S \ll N$. In this limit, we find

\begin{eqnarray}
\langle \hat{N}_0 \rangle_{SM} &  \approx (N^2-S^2)(S^2-M^2)/8,\\
\langle \hat{N}_0^2 \rangle_{SM}  &  \approx (3N^2-S^2)(S^2-M^2)^2/8,\\
\left( \Delta N_0^2\right)_{SM} &  \approx (N^2-S^2)(S^2-M^2)^2/8.
\end{eqnarray}
where $\langle \hat{N}_0^p \rangle_{SM} = \langle N,S,M \mid N_{0}^p \mid N,S,M \rangle$. Hence, we find super-Poissonian fluctuations for $M \ll S \ll N$, which eventually vanish as $S$ (resp. $M$) increases to its maximum value $N$ (resp. $S$).

We calculate now the thermally averaged $\langle n_{0}\rangle_{T}$ and $\left( \Delta n_{0}^2\right)_{T}$ in the canonical ensemble. The average population in $m=0$ is given by
\begin{eqnarray}
\langle N_{0}\rangle_{T} & = \frac{1}{Z} \sum_{S,M}e^{-\beta' S(S+1)} \langle  N_{0}\rangle_{SM}.
\end{eqnarray}
The second moment $\langle N_{0} ^2\rangle_{T} $ and the variance $\left( \Delta N_{0}^2 \right)_T$ are given by similar expressions. Here $Z$ is the partition function and $\beta'=U_{s}/2Nk_{B}T$. Assuming that the temperature is large compared to the level spacing ($k_{B}T \gg U_{s}/N$), the thermodynamic sums over energy levels is dominated by states with large $S\gg1$. There are two regimes to consider. 

At intermediate temperatures, states with $1 \ll S \ll N$ dominate the thermodynamics. To calculate the thermal average over all $S$ in this regime, we replace the discrete sums by integrals and send the upper bound $N$ of the integral to infinity. A simple estimate of the mean value of $S$,  $\langle S \rangle \sim (N k_{B}T/U_{s})^{1/2}$, shows that the condition $1 \ll S \ll N$ corresponds to the boundaries 
\begin{eqnarray}\label{eq:domain}
\frac{U_s}{N} \ll k_{B}T \ll N U_s.
\end{eqnarray}
In this regime, we find
\begin{eqnarray}
\label{FlucN0Ta}
\langle N_{0}\rangle_{T} & \approx \frac{N}{3},   \\
\label{FlucN0Tb}
\left( \Delta N_{0}^2 \right)_{T } & =\langle N_{0}^2\rangle_{T}-\langle N_{0}\rangle_{T}^2  \approx \frac{4 N^2}{45},~~k_B T \ll NU_s.
\end{eqnarray}
We note that to leading order in $1/N$, the moments of $N_0$ are identical for those found in the singlet state.

The second regime arises when the temperature becomes very large ($k_{B}T/N U_{s}> 1$), where one expects the sum to saturate due to the finite number of states. In this limit, the upper bound of the integral cannot be taken to infinity, and one must take the restriction $S\leq N$ into account. On the other hand, the Boltzmann factor can be replaced by unity, and the sums can then be calculated analytically. One finds

\begin{eqnarray}
\left( \Delta N_{0}^2\right)_{T\gg NU_s} & \approx N^2/18.
\end{eqnarray}

To summarize (see Figure~\ref{Fig:summary}), for $q=0$ we always find large depletion and super-Poissonian fluctuations ($\Delta N_{0}^2\sim \langle N_{0}\rangle ^2$). The average population is always $N/3$ as expected from the isotropy of the Hamiltonian. The relative standard deviation remains approximately constant (to order $N$) at the value $ \Delta N_{0}/N\approx 2/3\sqrt{5}\approx 0.298$ for $k_B T \ll NU_s$, and changes to $1/3\sqrt{2}\approx 0.236$ for very large temperatures $k_B T > NU_s$ where all states are occupied with equal probability.

\subsection{Bogoliubov approximation for $q\neq0$}
\label{sec:spinfragbogo}
 For large $q>0$ (and $\langle S_z \rangle$ constrained to vanish only in average), we expect that the system will form a condensate in the $m=0$ Zeeman state, with small fluctuations. Such a system can be described in the Bogoliubov approximation (as described in the Appendix of \cite{barnett2011a}), which extends to any $M$ the harmonic oscillator approximation made earlier for the $M=0$ sector. One sets ${\hat a}_{0} \approx  \sqrt{N}_{0}$, and expresses the fluctuations ${\hat a}_{\pm 1}$ in terms of new operators $\hat{\alpha}^\pm$,
 \begin{eqnarray}
\hat{\alpha}_{\pm} & = & u {\hat a}_{\pm 1}-v  {\hat a}_{\mp 1}^\dagger.
\end{eqnarray}
Here the Bogoliubov amplitudes $u,v$ defined by
\begin{eqnarray}
u\pm v & = &  \left( \frac{q}{2U_s+q}\right)^{\pm 1/4}.
\end{eqnarray}
 are chosen to put the Hamiltonian in diagonal form,
\begin{eqnarray}
H_{\rm Bogo} &=&\sum_{\mu=\pm} \hbar \omega \left(  \hat{\alpha}_{\mu}^\dagger \hat{\alpha}_{\mu} +\frac{ 1}{2}\right)-(g+q).
\end{eqnarray}
The energy $\hbar \omega$ of the Bogoliubov mode is identical to the one previously found in the harmonic oscillator approximation for $M=0$ [Eq.~(\ref{eq:bogofreq})]. Note that we have now two such modes (instead of only one in the case $M=0$)\footnote{We expect in general three modes of excitations for a spin $1$ system. When the constraint of constant particle number is taken into account, this reduces the number of modes to two. The suppressed mode would correspond to density fluctuations in an extended system, and is explicitly ruled out by the SMA. When a further constraint $M=0$ is imposed, another mode is cancelled - corresponding to magnetization fluctuations which are explicitly forbidden, thus leaving only one excitation mode.}.

In the Bogoliubov approximation, the moments of $N_0$ can be obtained analytically. The quantum ($T=0$) depletion of $N_0$ is smaller than one atom. The thermal part of the depletion and variance of $n_0=N_0/N$ read for  $k_{B}T\gg \hbar \omega$
\begin{eqnarray}
\label{eq:bogo1}
1-\langle n_0\rangle  &= \frac{2(U_s+q)}{q+2U_s} \frac{k_B T}{Nq},\\ 
\label{eq:bogo2}
\Delta n_0^2  &= \frac{2\left[(U_s+q)^2+U_s^2\right]}{(q+2U_s)^2} \left(\frac{k_B T}{Nq}\right)^2.
\end{eqnarray}
The prefactors take values of order unity, and both the depletion $1-\langle n_0\rangle$ and standard deviation $\Delta n_0$ scale as $k_B T /Nq$.
The above expressions are valid provided they describe small corrections to a regular polar condensate where almost all atoms accumulate in $m=0$ ($\langle n_0\rangle =1$), or in other words for temperatures
\begin{eqnarray}
k_B T \ll Nq.
\end{eqnarray}

\subsection{Comparison between the different approximations }
\begin{figure}[ht!!!!!]
\begin{center}
\includegraphics[width=17cm]{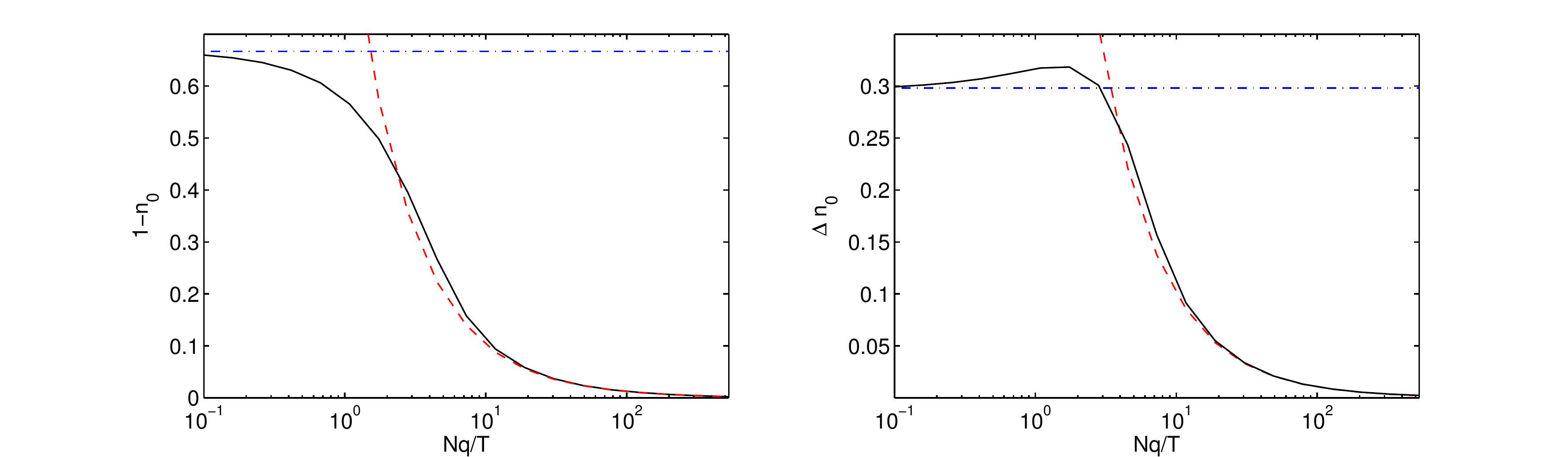}
\caption{Depletion (top) and standard deviation (bottom) of $N_0$. The solid line shows the exact numerical result for $N=1000$ and $T=10 U_s$, the blue dash-dotted line is the result calculated for $q=0$ using Eqs.~(\ref{FlucN0Ta},\ref{FlucN0Tb}), and the red dashed line shows the Bogoliubov approximation. Deviations are observed for $q/T\sim N$, which is expected from our approximation: This regime corresponds to a depletion of one atom or less, and corrections $\propto 1/N$ that we neglect become important.}
\label{Fig:SMA_M}
\end{center}
\end{figure}

We compare in Figure~\ref{Fig:SMA_M} the predictions for the moments of $N_0$ obtained from the various approximations discussed the paper, Bogoliubov approximation, and $q=0$ limit. These approximations are compared to the results obtained by diagonalization of the original Hamiltonian (\ref{eq:H}) and computing thermodynamic averages using the exact spectrum and eigenstates. 

When $Nq / k_B T \ll 1$, the localized states of Section \ref{sec:excitedM0}, which are dominated by their potential energy, will be populated. Because these localized states are close to the angular momentum eigenstates found in the $q=0$ limit, to a good approximation the formula derived in Section \ref{sec:spinfragq0} [see (\ref{FlucN0Ta},\ref{FlucN0Tb}) and the continous blue line in Figure \ref{Fig:SMA_M}]. On the other hand, for $Nq/k_B T \gg 1$, thermal states mostly populate states with $E \sim qN$, {\it i.e.} ``delocalized" states within the low-energy ``Bloch band" of width $\sim Nq$. Those states correspond to small depletion and fluctuations, and they are well described by the Bogoliubov approximation presented in Section \ref{sec:spinfragbogo} [see (\ref{eq:bogo1},\ref{eq:bogo2}) and the red dashed line in Figure \ref{Fig:SMA_M}]. The numerical solution of the original model (\ref{eq:matrix_equation}) interpolates between the two well-defined asymptotic limits, either a thermal mixture of total spin eigenstates for $q \ll k_B T /N$ or a thermal state of Bogoliubov-like excitations for $q \gg k_B T /N$.  

We note to conclude this section that in the regime $U_s/N \ll k_B T \ll N U_s$, the tight-binding model defined Eq.~(\ref{eq:matrix_equation}) has a quasi-universal form at finite temperatures, in the sense that the model is entirely specified by two dimensionless parameters, for instance $k_B T/U_s$ and $N q/U_s$. We found that the physical quantities $\langle N_0 \rangle, \left(\Delta N_0 \right)$ depend only on their ratio $Nq/k_B T$, to a very good approximation.  This quasi-universality, which can be explored by experiments, will be easily justified in the broken symmetry approach presented in the next Section.

\section{Comparison with the broken-symmetry picture}\label{sec:broken}

So far, we have treated the problem by the most natural method, by looking for the eigenspectrum of the Hamiltonian. Another approach \cite{castin2001a,mueller2006a} to the problem of spin 1 bosons with antiferromagnetic interactions relies on the set of so-called polar or spin-nematic states, defined as

\begin{eqnarray}
\vert N: \bf \Omega \rangle &= \frac{1}{\sqrt{N!}} \left( {\bf \Omega} \cdot {\bf \hat{a}} \right)^N \vert {\rm vac }\rangle,
\end{eqnarray}
where the vector $ {\bm \Omega} $ reads in the standard basis
\begin{eqnarray}
\label{eq:nematic}
 {\bm \Omega} =e^{i\gamma}\left( \begin{array}{c}
\frac{1}{\sqrt{2}}\sin(\theta)e^{i\phi} \\
\cos(\theta) \\
-\frac{1}{\sqrt{2}}\sin(\theta)e^{-i\phi} \end{array}\right).
\end{eqnarray}
For a single particle, the states $\vert  {\bm \Omega}\rangle = \sum_{i=0,\pm1} \Omega_i  \vert m=i\rangle$ form a continuous family of spin 1 wavefunctions with vanishing average spin. In fact, $\vert {\bm \Omega} \rangle$ is the eigenvector with zero eigenvalue of the operator ${\bm \Omega} \cdot {\bf \hat{s}}$, with ${\bf \hat{s}}$ the spin 1 operator. The states $\vert N: {\bf \Omega} \rangle$ correspond to a many-body wave function where all particle occupy the single-particle state $ \vert {\bm \Omega}\rangle$. As a result, one has $\langle N: {\bf \Omega} \vert {\bf \hat{S} }\vert N: {\bf \Omega} \rangle=0$. 

\subsection{Zero temperature}

 It is interesting to connect the spin nematic states to the angular momentum eigenstates. The spin nematic states form an overcomplete basis of the bosonic Hilbert space. Writing the states $\vert N, S, M\rangle$ in this basis, one finds \cite{castin2001a,mueller2006a,barnett2011a}
\begin{eqnarray}
\vert N, S, M\rangle = \int d{\bf \Omega} ~Y_{S,M} ({\bf \Omega}) \vert N: {\bf \Omega}    \rangle
\end{eqnarray}
where $Y_{SM}$ denotes the usual spherical harmonics and where $d{\bf \Omega}=\sin(\theta)d\theta d\phi$. In particular, the singlet ground state $\vert N, 0, 0 \rangle$ appears to be a coherent superposition with equal weights of the nematic states. Consider now the average value in the singlet state $\langle \hat{O}_k  \rangle_{\rm singlet} =\langle N,0,0 \vert \hat{O}_k \vert N,0,0 \rangle$of a $k-$body operator $\hat{O}_k$,
\begin{eqnarray}
\langle \hat{O}_k  \rangle_{\rm singlet}  = \frac{1}{4\pi} \int d{\bf \Omega} d{\bf \Omega'}  \langle N: {\bf \Omega'} \vert \hat{O}_k \vert N: {\bf \Omega}\rangle. 
\end{eqnarray}
As shown in \cite{castin2001a}, for few-body operators with $k \ll N$ this expectation value can be approximated to order $1/N$ by the much simpler expression
\begin{eqnarray}
\langle \hat{O}_k  \rangle_{\rm singlet}   \approx \frac{1}{4\pi} \int d{\bf \Omega}  \langle N: {\bf \Omega} \vert \hat{O}_k \vert N: {\bf \Omega}\rangle. 
\end{eqnarray}


This approximation shows that the system can equally well be described by a statistical mixture of spin-nematic states described by the density matrix \cite{castin2001a,mueller2006a}
\begin{eqnarray}
\hat{\rho}_{BS} = \frac{1}{4\pi} \int d{\bf \Omega} \vert N: {\bf \Omega} \rangle  \langle N: {\bf \Omega} \vert.
\end{eqnarray}
At zero temperature and zero field, there is no preferred direction for the vector ${\bf \Omega}$ so that each state can appear with equal probability. This approach is known as a ``broken symmetry" point of view, where one can imagine that the atoms condense in the same spin state for each realization of the experiment, but this spin state fluctuates arbitrarily from one realization to the next. The important point is that the overlap integral $\langle N: {\bf \Omega'} \vert N: {\bf \Omega}\rangle$ between two spin-nematic states vanishes very quickly with the distance $\vert \bf \Omega-\bf \Omega'\vert$. This allows one to use the approximation $\langle N: {\bf \Omega'} \vert N: {\bf \Omega}\rangle \approx \delta({\bf \Omega-\bf \Omega'})+ {\cal O}(1/N)$, which leads to
\begin{eqnarray}
\langle \hat{O}_k  \rangle_{\rm BS}  = {\rm Tr}\left[ \hat{\rho}_{BS} \hat{O}_k\right] \approx \langle \hat{O}_k  \rangle_{\rm singlet}  + {\cal O}(1/N).
\end{eqnarray}
This result can be written as a general statement concerning average values of few-body observables with $k\ll N$ \cite{castin2001a}: to leading order in $1/N$, the exact and broken symmetry approaches will give the same results after averaging over the ensemble. The differences between the two approaches are subtle and vanish in the thermodynamic limit as $1/N$. 

It is worth noting the difference between individual states and the ensemble. The moments of $N_0$ in the state $\vert N:{\bf \Omega} \rangle$ are given by
\begin{eqnarray}\nonumber 
\langle N:{\bf \Omega}\vert \hat{N}_0 \vert N:{\bf \Omega} \rangle &= &N u^2,\\ \nonumber 
\langle N:{\bf \Omega}\vert \hat{N}_0^2 \vert N:{\bf \Omega} \rangle &=& N(N-1)u^4+N u^2,
\end{eqnarray}
where $u=\cos(\theta)$. The variance of $N_0$ for a system prepared in a single spin-nematic state, $N u^2(1-u^2)$, is thus Poissonian, as expected for a regular condensate. On the other hand, computing the ensemble averages over $\hat{\rho}_{\rm BS}$ gives

\begin{eqnarray}
\langle N_0 \rangle &=&N \int_0^1 du \,u^2 = \frac{N}{3}, \nonumber  \\
\langle N_0^2 \rangle &=& \int_0^1 du  \left( N(N-1) u^4 +N u^2 \right) \nonumber  \\
&=&\frac{3N^2+2N}{15},\nonumber \\
\Delta N_0^2&=&\frac{4N^2+6N}{45}.\nonumber 
\end{eqnarray}

The variance in the ensemble is thus super-Poissonian, and differs from the result in the exact ground state only by the sub-leading term $\propto N$. This is in agreement with the general statement made above.

\subsection{Moments of $N_0$ at finite temperatures}

We now extend the broken symmetry approach summarized above to finite temperatures. The density matrix should include a weight factor proportional to the energy of the states $\vert N:\Omega \rangle$. To leading order in $1/N$, these states have zero interaction energy \footnote{Explicitely, one has $\langle N: {\bf \Omega} \vert {\bf \hat{S} }^2 \vert N: {\bf \Omega} \rangle=2N \cos^2(\theta)$, so that the interaction energy of the state $\vert N:{\bf \Omega}\rangle$ is given by $U_s\cos^2(\theta)\sim \mathcal{O}(1)$ compared to the QZ energy $\sim\mathcal{O}(N)$. The same argument applies to off-diagonal matrix elements $\langle N: {\bf \Omega}' \vert {\bf \hat{S} }^2 \vert N: {\bf \Omega} \rangle$. } and a mean QZ energy given by $-N q \cos^2(\theta)$. In the spirit of the mean-field approximation, we replace the Boltzmann factor by its mean value and write the density matrix as

\begin{eqnarray}
\hat{\rho}_{BS} 
\approx \frac{1}{{\cal Z} } \int d{\bf \Omega} \vert N: {\bf \Omega} \rangle  \langle N: {\bf \Omega} \vert e^{N \beta q \Omega_z^2},\label{eq:dm}
\end{eqnarray}
with $\beta =1/k_B T$. The partition function can then be expressed as
\begin{eqnarray}\nonumber
{\cal Z} &=& \int_{0}^{2\pi}d\phi\int_{0}^{\pi} \sin(\theta)d\theta~  e^{N \beta q \cos^2(\theta)} \\
&=&4\pi F_{-1/2}\left(N\beta q\right).
\end{eqnarray}
Here we introduced the family of functions
\begin{eqnarray}
F_\alpha\left(  y\right)=\int_0^y x^{\alpha-1} e^{-x} dx
\end{eqnarray}
which are related to the lower incomplete gamma functions.
In a similar way, we can compute the moments of $n_0=N_0/N$ to leading order in $N$ as
\begin{eqnarray}
\langle n_0^m \rangle &= \frac{F_{m-1/2}\left( N\beta q\right)}{F_{-1/2} \left( N\beta q\right)}.
\end{eqnarray}
From this result, one can easily deduce the average and variance of $n_0$. This calculation provides an explicit proof of the numerical evidence that, to leading order in $N$, the moments of $N_0$ obey a universal curve depending only on $Nq/k_B T$ and not on $q/U_s$ or $T/U_s$ separately. 

From the properties of the functions $F_\alpha$, we recover the results established in the previous Section. When $x\rightarrow0$, $F_\alpha (x) \sim x^\alpha/\alpha$. Using this result we recover for $q =0$ the previous results, {\it i.e.} $\langle n_0 \rangle=1/3$, $\langle n_0^2 \rangle=1/5$ and $\Delta n_0^2=4/45$. When $x\rightarrow \infty$, $F_\alpha (x) \sim e^x/x\times \left [1-\alpha/x+(\alpha-1/2)(\alpha-3/2)/x^2\right]$. This leads to the asymptotic behavior $\langle n_0^m \rangle \sim 1-m/(N \beta q)+ (m^2-3m/2)/(N \beta q)^2 +\cdots $ when $N \beta q \gg 1$, which reproduces the Bogoliubov results (\ref{eq:bogo1},\ref{eq:bogo2}) for $q\ll U_s$.

\begin{figure}[h!!!!!!]
\begin{center}
\includegraphics[width=17cm]{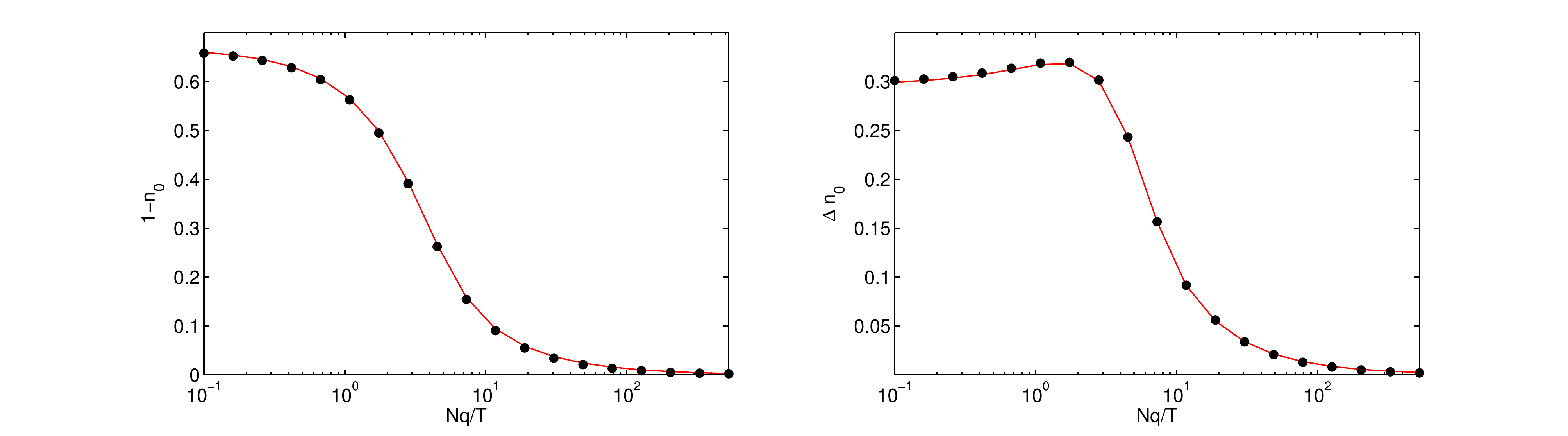}
\caption{Exact diagonalization (red solid line) vs broken symmetry approach (black dots) for $N=1000$, $k_B T/U_s=10$.}
\label{Fig:SMA_M_broken}
\end{center}
\end{figure}

We finally compare in Figure~\ref{Fig:SMA_M_broken} the results from the broken symmetry approach to the results obtained by diagonalizing the Hamiltonian \ref{eq:H}. We find excellent agreement between the two in the regime of thermal fragmentation, supporting the picture of mean-field states with random orientation fluctuating from one realization to the next. We note that the {\it ansatz} (\ref{eq:dm}) for the density matrix  is by no means obvious, and the good agreement with the numerical results is obtained only because the set of polar states is a good description for sufficiently low temperatures : Although these states are not true eigenstates of the Hamiltonian (\ref{eq:H}), the action of $\hat{H}$ yields off-diagonal matrix elements scaling as $1/N$ \cite{anderson1952a}, and thus vanishing in the thermodynamic limit. At high temperatures ($k_BT \sim N U_s$), where all high energy states are populated the broken-symmetry ansatz is no longer adequate.

\section{Conclusion}
\label{sec:conclusion}

We have studied the properties of an ensemble of antiferromagnetic spin 1 bosons with QZ energy breaking the spin rotational symmetry. The system evolves with increasing QZ energy from a super fragmented condensate with large fluctuations to a regular polar condensate where atoms condense in $m=0$. We focused in particular on the behavior of a thermal mixture of excited states, and discussed the evolution of the moments of $N_0$ with increasing $q$. Two approaches were explored, one relying on diagonalization of the Hamiltonian (either exactly or approximately in certain parameter regimes), and the other relying on a broken symmetry picture where the system is described as a statistical mixture of degenerate polar condensates. Both approaches were found in remarkable agreement. In this article, we focused on equilibrium properties and assumed thermal equilibrium from the start. An interesting question is how the physical system ({\it i.e.} also including the dynamics of non-condensed modes not described in the SMA) can reach such an equilibrium state, {\it e.g.} following a quench in $q$ \cite{pu1999a}. This problem, which can be linked to the more general question of thermalization of closed quantum systems \cite{polkovnikov2011a} provides an interesting direction for future work.

\appendix
\section{Total spin eigenstates}\label{app:wf}

The general expression of the states $\vert N,S,M \rangle$ in the Fock basis is
\begin{eqnarray}
\vert N,S, M \rangle = \frac{1}{\sqrt{\mathcal{N}}} \left( \hat{S}^{(-)}\right)^P \left( \hat{A}^\dagger \right)^Q \left( \hat{a}_{+1}\dagger\right)^S \vert {\rm vac} \rangle.
 \end{eqnarray}
Here $P=S-M$, $2Q=N-S$, $\hat{S}_{-}$ is the spin lowering operator and $\hat{A}^\dagger  = \hat{a}_{0}^\dagger-2  \hat{a}_{-1}^\dagger  \hat{a}_{+1}^\dagger$is the singlet creation operator. The two operators commute. The normalization constant reads
\begin{eqnarray}
\mathcal{N} & =  \frac{S ! (N-S)!! (N+S+1)!! (S-M)! (2S)!}{(2S+1)!! (S+M)!},
 \end{eqnarray}
where $!!$ indicates a double factorial. 

The action of $\hat{a}_{0}$ on the angular momentum eigenstates is 
\begin{eqnarray}\nonumber
\hat{a}_{0}\vert N,S,M\rangle & =\sqrt{A_{-}(N,S,M)} \vert N-1,S-1,M\rangle\\ &  + \sqrt{A_{+}(N,S,M)} \vert N-1,S+1,M\rangle,
\label{eq:a0}
\end{eqnarray}
where $\hat{a}_{0}$ is the annihilation operator of a boson in the Zeeman state $m=0$, and where the coefficients $A_\pm$ are given by
\begin{eqnarray}
A_{-}(N,S,M) & = \frac{(S^2-M^2)(N+S+1)}{(2S-1)(2S+1)},\\
A_{+}(N,S,M) & =  \frac{((S+1)^2-M^2)(N-S)}{(2S+1)(2S+3)}.
\end{eqnarray}
The non-zero matrix elements of $\hat{N}_{0}$ are

\begin{eqnarray}
 \langle S\vert \hat{N}_{0}\vert S \rangle &= \left(A_+(N,S,M)
+ A_-(N,S,M)\right),   \\
 \langle S+2\vert \hat{N}_{0}\vert S \rangle &=  \sqrt{A_-(N,S+2,M)A_+(N,S,M) }, \\
  \langle S-2\vert \hat{N}_{0}\vert S \rangle &= \sqrt{A_+(N,S-2,M)A_-(N,S,M) }.
\end{eqnarray}
where we abbreviated the notation for the state $\vert N,S,M\rangle$ as $\vert S\rangle$ to simplify the notation.
We then obtain the matrix elements of $\hat{H}_0$ in the $\vert N,S,M \rangle$ basis as
\begin{eqnarray}
h_{S,S}^M  &= \frac{U_s}{2N}S(S+1) -q \langle S\vert \hat{N}_{0}\vert S \rangle,\\
h_{S,S+2}^M  &=  -q  \langle S+2\vert \hat{N}_{0}\vert S \rangle ,\\
h_{S,S-2}^M  &=  -q   \langle S-2\vert \hat{N}_{0}\vert S \rangle.
\end{eqnarray}

\section{Continuum approximation}\label{app:TB}
We expand the matrix elements $h_{S,S}, h_{S,S \pm 2}$ to first order in $1/S, S/N, M^2/S^2$, and obtain
\begin{eqnarray}
h_{S,S\pm 2}^M &\approx \frac{N}{4}\left[1-\frac{M^2}{(S\pm1)^2}\right] -\frac{1}{8N}\left[ (S\pm1)^2-M^2 \right],\\
h_{S,S}^{M \neq 0}  & \approx \frac{N}{2}\left(1-\frac{M^2}{S^2}\right),~~ h_{S,S}^{M = 0}   \approx \frac{N}{2}.
\end{eqnarray}
For $M=0$, we obtain
\begin{eqnarray}
\nonumber
-\frac{Nq}{4}\left[\left(1-\frac{(x+\epsilon)^2}{2}\right) c_{S+2}+\left(1-\frac{(x-\epsilon)^2}{2}\right) c_{S-2}\right]\\
+\frac{NU_s}{2} x^2 c_S=\left(E +\frac{Nq}{2}\right) c_S,
\end{eqnarray}
where we have set $x=S/N$ and $\epsilon=2/N$.  We now take the continuum limit, where $\epsilon \ll 1$ is taken as a discretization step  and $c_s$ becomes a continuous function $c(S)$. We write
\begin{eqnarray}
\frac{N^2}{4} \left(c_{S+2}+c_{S-2} \right) \approx \Delta c(s) +\frac{N^2}{2} c(s). 
\label{eq:TB2}
\end{eqnarray}
Substituting in (\ref{eq:TB2}) and neglecting a term $\propto (q x^2/N)\Delta c$, we arrive at (\ref{eq:HO}). 

This derivation is valid as long as the relevant states are well localized around $x=0$. This is always the case in the ground state, which has a width at most $\sim 1/\sqrt{N}$ for $q\gg U_s$. For thermal states, the width is $\sim \sqrt{k_B T/\left[N(2U_s+q)\right]}$, which gives the condition $k_B T \ll N(U_s+q)$. Finally, the cross-term $\propto (q x^2/N)\Delta c$ is of order $2E_p E_c c /\left [ N(2U_s+q)\right]$ in terms of the kinetic and potential energies $E_c ,E_p$ of the harmonic oscillator. In the thermal regime, a typical order of magnitude for this term is thus $(k_BT)^2/[N(2U_s+q)]$, small compared to the energy $k_B T$ typical for the other terms we kept in the equation provided the condition above is fulfilled.

We acknowledge discussions with members of the LKB, in particular Yvan Castin. This work was supported by IFRAF, by Ville de Paris (Emergences project) and by DARPA (OLE project).

\section*{References}
\bibliographystyle{unsrt}

\bibliography{BibSpinor}

%
%
%
%

\end{document}